\documentstyle[11pt,IAUS212,twoside,epsf]{article}

\def\edcomment#1{\iffalse\marginpar{\raggedright\sl#1\/}\else\relax\fi}
\reversemarginpar

\begin{document}
\vspace*{1cm}
\title{A Near-Infrared Survey for Galactic Wolf-Rayet Stars}

\author{Nicole Homeier}
\affil{ESO-Garching, Karl-Schwarzschild Str. 2, D-85748, Garching, Germany}

\author{Robert Blum}
\affil{Cerro Tololo Interamerican University, La Serena, Chile}

\author{Peter Conti}
\affil{JILA, U Colorado, Boulder, CO}

\author{Anna Pasquali}
\affil{ST-ECF/ESO-Garching, Germany}

\author{Augusto Damineli}
\affil{IAG, S\~{a}o Paulo, Brazil}

\begin{abstract}

Most of the Milky Way's evolved massive stellar population is hidden from 
view. We can attempt to remedy this situation with near-infrared 
observations, and in this paper we present our method for detecting 
Wolf-Rayet stars in highly extincted regions and apply it to the 
inner Galaxy. Using narrow band filters at K-band wavelengths, we 
demonstrate how WR stars can be detected 
in regions where they are optically obscured. Candidates are selected
for spectroscopic follow-up from our relative line and continuum 
photometry. The final results of applying this method with a NIR survey 
in the Galactic plane will provide a more complete 
knowledge of the structure of the galactic disk, the role of metallicity 
in massive stellar evolution, and environments of massive star formation.
In this paper we briefly describe the survey set-up and report on 
recent progress. We have discovered four emission-line objects in the inner 
Galaxy: two with nebular emission lines, and two new WR stars, both of 
late WC subtype.

\end{abstract}

\section{Introduction}

Optical surveys within our Galaxy are severely hampered by dust obscuration;
therefore complete samples must be obtained with 
longer wavelength observations. Here we describe a survey in the Galactic
plane at K-band wavelengths, where the extinction is much lower than
traditional V-band surveys. Our scientific driver is the discovery of the
young stellar population in our Galaxy through the detection of evolved
massive stars. These stars have strong emission lines, which makes them
relatively easy to detect using narrow-band filters.

As a massive O star evolves, its spectrum becomes dominated by emission
lines, arising either in a dense stellar wind, or in circumstellar
material produced by mass loss. Stellar emission lines are most pronounced 
in Wolf-Rayet (WR) stars which have lifetimes $<$~10~Myr, and thus are 
excellent tracers of recent star formation, and so also Galactic structure. 
A complete sample of Galactic WR stars will also enable us to understand 
the distribution of WN and WC sub-types in a high metallicity environment.
Additionally, WR stars are critical components
in our quest to understand how star formation proceeds. For example, most
of the previously known WRs are relatively isolated, but
recent searches in the IR have found a plethora of these objects in
stellar clusters near the Galactic center (e.g. Figer et al. 1999). 

\section{Survey Description}

We employ four filters situated on prominent
lines of He I (2.06 um), C IV (2.08 um), Br $\gamma$ (2.166 um), 
He II (2.19 um),
and three continuum filters covering wavelengths bluer and redder than
each of the line filters (2.03, 2.14, and 2.248 um) (Blum \& Damineli 1999,
Homeier et al. 2000). In this way we minimize
the effects of differential reddening on the continuum of each individual 
star. Each frame is analyzed with the set of routines collectively called the 
'DoPhot' package, a psf-fitting photometry package optimized for large
images done in a pipeline process. DoPhot identifies,
classifies, and performs photometry on objects in an image, making
successive passes over the image to search each time for fainter objects.
Candidates are selected for spectroscopic follow-up based on 
relative line and continuum photometry and image subtraction with 
visual examination of the output (Homeier et al. 2002, in prep).

In 1996 we began taking data at the 1.5m at CTIO, using the Cerro Tololo 
Infrared Imager (CIRIM). This data comprises approximately 1.5 sq degrees 
of sky near the Galactic Center, half of which we have follow-up 
spectroscopic observations for. With this data we have discovered 
two new WR stars and two objects with surrounding nebular emission. 

\section{Newly Discovered Emission-line Objects}

\begin{table}
\caption[]{}
\begin{tabular}{lllc}\hline
Star 		   & RA (J2000) & Dec (J2000) & J, H, K \\\hline
WR1  		   & 17:45:42.3 & -28:52:54.7 & NA, NA, 10.5?\\	
WR2 (Quin. source) & 17:46:13.2 & -28:49:30.1 & 13.3, 11.0, 9.4 (fm. 2MASS)\\
\hline
\end{tabular}
\end{table}

\begin{figure}
\plottwo{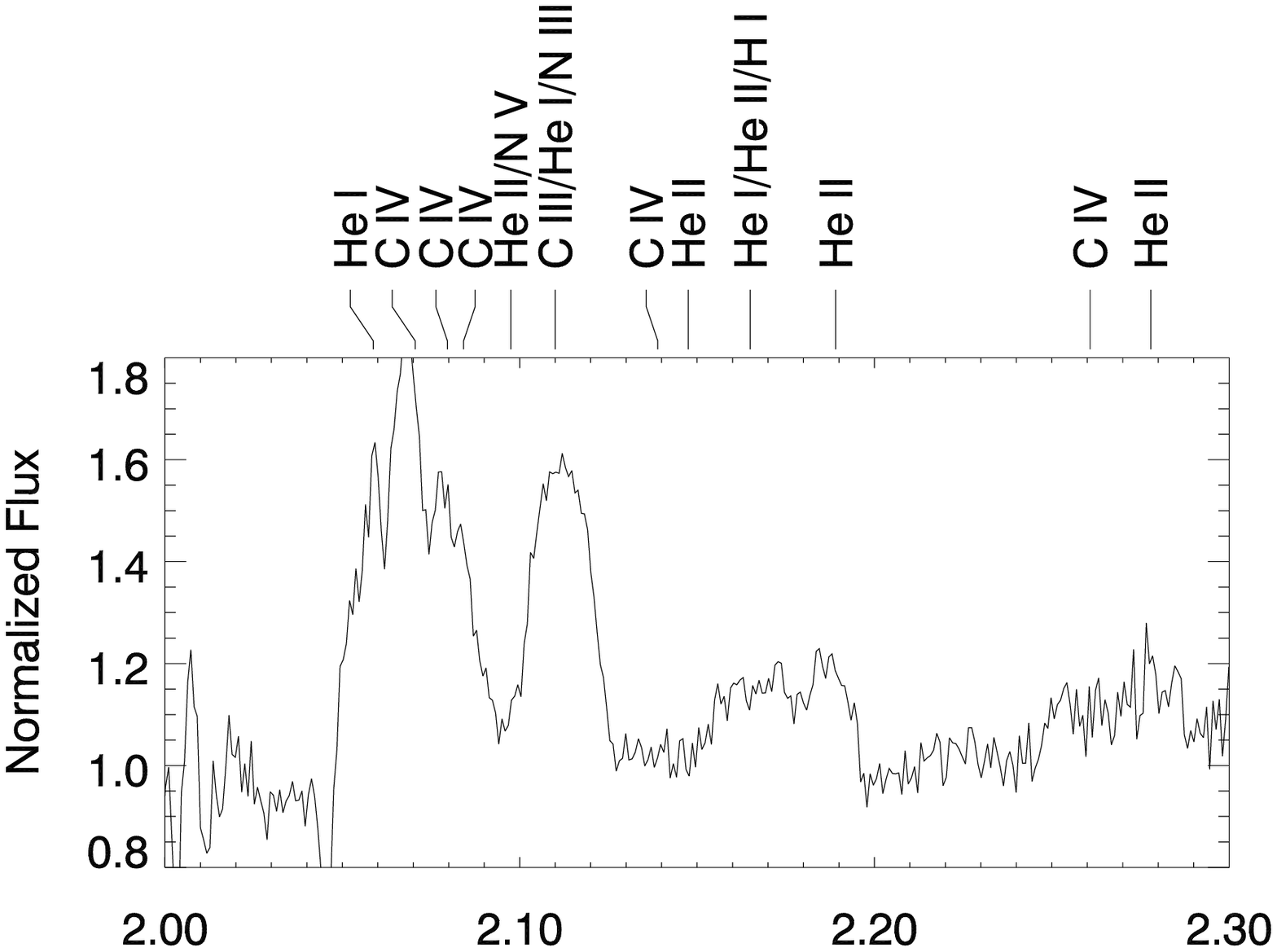}{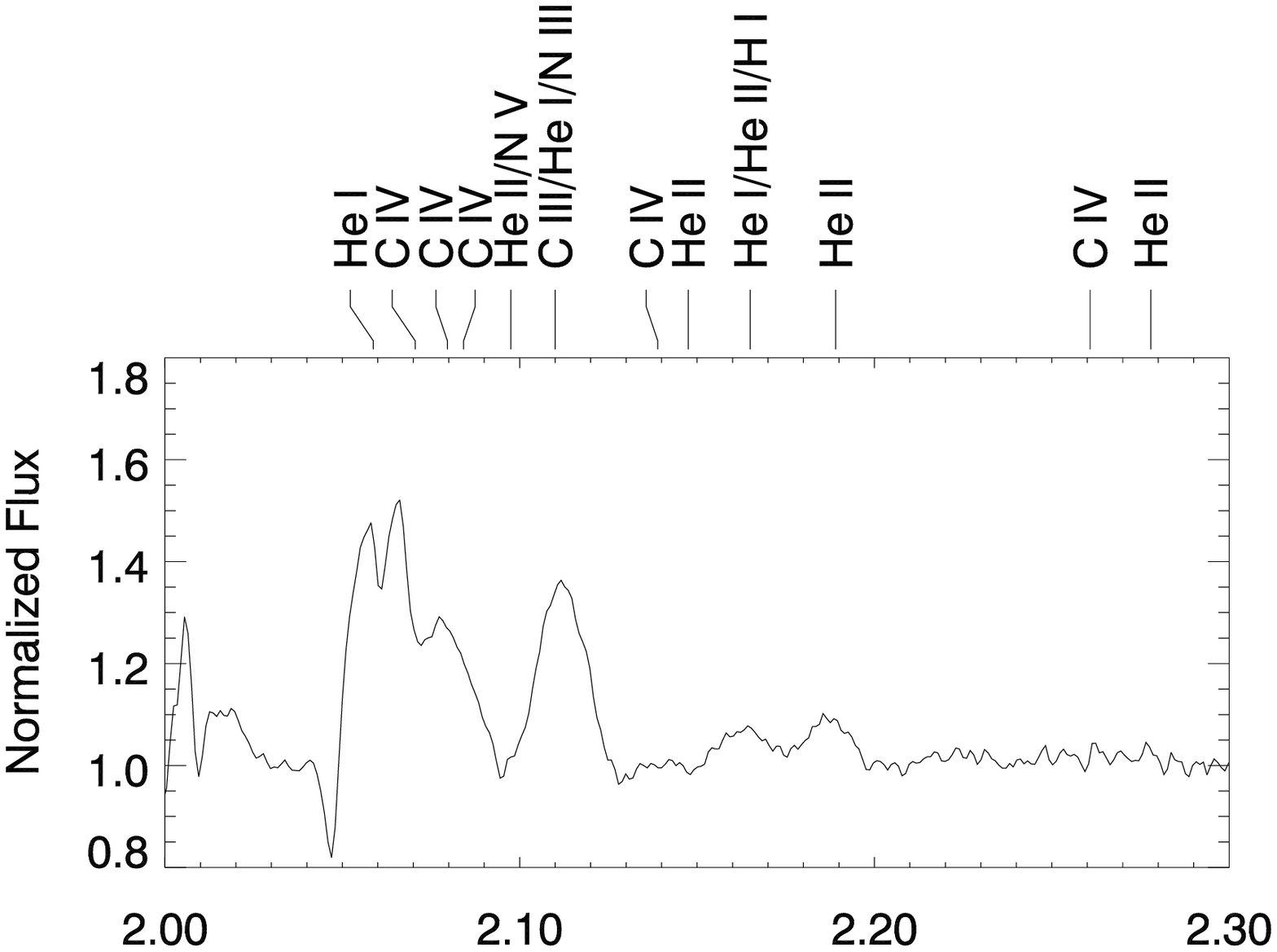}
\caption{K-band spectra of the two newly discovered WR stars in our survey.
Both have K-band spectra indicative of a late WC type, WC7-8.
To the left is the star which does not currently belong to a cluster, and on
the right is the WR discovered in the Quintuplet cluster.}
\label{spectra}
\end{figure}

Both of the newly discovered WRs appear to be late-type WC stars (WC7-8).
One is $\sim 8$~arcmin from the Quintuplet cluster (WR1), or 20~pc projected 
distance. We speculate that it has been ejected from the Quintuplet
or is a former member of an as-yet undetected massive stellar cluster
in the Galactic Center region. The second WR is part of the Quintuplet 
(WR2, Quintuplet source), 
and is located just outside the region examined by Figer et al. (1999). 
Spectra of these two objects are presented in Figure 1, and coordinates are
listed in Table 1. Finding charts will be published in an upcoming paper
(Homeier et al. 2002, in prep).

Of the two objects with nebular emission, one is surrounded by a spherical
nebula with strong H I lines, including Br $\gamma$ 2.166 $\mu$m, and 
He I 2.06 $\mu$m emission. This object could be a PN, LBV, or a young O 
star ionizing the surrounding gas. It is located on the edge of 
a region with extremely large extinction, and there
appears to be more nebular emission a few arcseconds to the south-east. We
speculate that this is a star-forming region on the edge of a molecular
cloud, and further observations are needed to deduce the distance of this 
region and whether it contains high-mass stars.


\begin{references}

\reference Blum, R., \& Damineli, A. 1999, in K. A. van der Hucht, 
G. Koenigsberger, \& P. R. J. Eenens (eds.), Wolf-Rayet Phenomena in
Massive Stars and Starburst Galaxies, Proc. IAU 193 (San Francisco: ASP), 472 

\reference Figer, D., McLean, I., \& Morris, M. 1999, ApJ, 514, 202

\reference Homeier, N., Blum, R., Conti, P., Damineli, A. 2000 AAS, 197.0412 

\reference Homeier, N., Blum, R., Conti, P., Damineli, A., Pasquali, A. 2002, in prep for A\&A

\end{references}
\end{document}